# Laser-induced periodic surface structured electrodes with 45 % energy saving in electrochemical fuel generation through field localization

*Chaudry Sajed Saraj*[1, 2], *Subhash C. Singh*[1 3*] *Gopal Verma*[1], *Rahul A Rajan*[1,2], *Wei Li,*[1,2*] *and Chunlei Guo*[3*]

[1]GPL, State Key Laboratory of Applied Optics, Changchun Institute of Optics, Fine Mechanics and Physics, Chinese Academy of Sciences, Changchun 130033, China
[2]University of Chinese Academy of Sciences (UCAS), Beijing 100049, China
[3]The Institute of Optics, University of Rochester, Rochester, NY 14627, USA
[*]Corresponding author: ssingh49@ur.rocheester.edu; weili1@ciomp.ac.cn; guo@optics.rochester.edu

**Abstract**

**Electrochemical oxidation/reduction of radicals is a green and environmentally friendly approach to generate fuels. These reactions, however, suffer from sluggish kinetics due to a low local concentration of radicals around the electrocatalyst. A large electrode potential can enhance the fuel generation efficiency via enhancing the radical concentration around the electrocatalyst sites, but this comes at the cost of electricity. Here, we report about 45 % saving in energy to achieve an electrochemical hydrogen generation rate of $3\times10^{16}$ molecules $cm^{-2}s^{-1}$ (current density: 10 mA/cm$^2$) through localized electric field-induced enhancement in the reagent concentration (LEFIRC) at laser-induced periodic surface structured (LIPSS) electrodes. The finite element model is used to simulate the spatial distribution of the electric field to understand the effects of LIPSS geometric parameters in field localization. When the LIPSS patterned electrodes are used as substrates to support Pt/C and RuO$_2$ electrocatalysts, the $\eta_{10}$ overpotentials for HER and OER are decreased by 40 and 25 %, respectively. Moreover, the capability of the LIPSS-patterned electrodes to operate at significantly reduced energy is also demonstrated in a range of electrolytes including alkaline, acidic, neutral, and seawater. Importantly, when two LIPSS patterned electrodes were assembled as the anode and cathode into a cell, it requires 330 mVs of lower electric potential with enhanced stability over a similar cell made of pristine electrodes to drive a current density of 10 mA/cm$^2$. This work demonstrates a physical and versatile approach of electrode surface patterning to boost electrocatalytic fuel generation performance and can be applied to any metal and semiconductor catalysts for a range of electrochemical reactions.**

# 1. Introduction

The incessantly growing global energy demands and greenhouse gas emissions pose enormous pressure on developed and developing countries to urgently replace their fossil-fuel-based conventional energy sources with renewables ones that have zero or negative carbon emissions.[1,2] Electrochemical (EC) and photoelectrochemical (PEC) oxidation/reduction of radicals at electrodes are green, environmentally friendly, and sustainable approaches, and are currently used in the generation of hydrogen, oxygen, ammonia, hydrocarbons, and other fuels.[3–6] These oxidation/reduction reactions, however, suffer from sluggish kinetics owing to the low local concentration of radicals close to the electrode surface. The concentration of radicals and bond activation, thus the rate of fuel generation, can be increased via the application of a large external electric potential, but this comes at the cost of consumed electricity.[7,8] An enormous amount of efforts have been devoted to designing and producing EC or PEC catalysts that could efficiently operate near-zero or at a low electric potential to develop self or atmospheric energy-powered fuel cells.[9–14] Most of these efforts, however, were centred on lowering the activation energy barrier i.e. thermodynamics of catalysts either via edge site engineering through nano-structuring [15–20] or enhancing the intrinsic activity of edge sites through chemical doping.[21–23] Little work has, however, been done in modulating local electric fields surrounding the catalysts that can directly govern the local concentration of radicals to enhance the kinetics of the reactions.

Recently, some efforts have been made to computationally optimize and experimentally realize the roles of long-range electric fields at the electrode-electrolyte interfaces or at the active sites to boost the biocatalytic performance of enzymes.[24,25] Sargent et al. demonstrated an enhancement in the electrocatalytic $CO_2$ reduction performance through modulating reagent concentration via electric field localization at gold tip electrodes, which shows that optimization of kinetics can be one of the most efficient ways to design high-performance electrocatalysts.[26,27] Orientation of external electric fields at the surface of a catalyst can be used to catalyze and control several non-redox, even nonpolar, reactions by several orders of magnitudes,[28,29] and to tune catalytic selectivity.[30] Very recently, Yu et al. demonstrated that the catalysts with ordered nanostructures are even better to regulate the surface flux of the reactant molecules for their proper utilization in the chemical reactions resulting in ~ 2 times higher current density,[31] which strongly indicates that the design and optimization of periodic surface patterns on catalysts surface can be one of the most efficient ways to realize next-generation smart catalysts.

In general, the fabrication of periodic nanopatterns on a metal surface relies either on multi-step, tedious and time-consuming lithography-based selective etching/deposition techniques or self-assembling routes[32,33]. Recently, we have demonstrated that femtosecond laser can create high-quality periodic patterns with sub-micron resolution, creating the so-called laser-induced periodic surface structures (LIPSSs. In contrast to complex lithography methods, LIPSSs can be produced in an ambient environment on a flat or curved surface of any materials such as metals, semiconductors, and dielectrics using a single step, faster, and mask less approach.[34–36] Surprisingly, LIPSSs have never been used for electric field localization to boost electrochemical fuel generation performance until this work.

Here, we fabricate LIPSSs on Ni foam (NF), one of the widely used electrode materials, and control geometric parameters to optimize localized electric field-induced modulation in the reagent concentration (LEFIRC) at the electrode surface. The LIPSSs patterned electrodes have hierarchal structures with periodic ridges and grooves of 100-300 nm widths covered with spherical nanoparticles (NPs) of 3-94 nm in diameters. The LEFIRC effect to boost the electrochemical fuel generation performance is tested for electrochemical hydrogen and oxygen generations through hydrogen evolution reaction (HER) and oxygen evolution reaction (OER), respectively. The optimized LIPSS patterned electrode demonstrates about 45 % reduction in the required electrical energy to achieve a hydrogen generation rate of $3\times10^{16}$ molecules cm$^{-2}$s$^{-1}$ (current density: 10 mA/cm$^2$). Furthermore, the applicability of the LIPSS patterned electrodes as a substrate to support and boost the performance of electrocatalyst powders through the LEFIRC effect is also examined. Pt/C, the model HER electrocatalyst, loaded on the LIPSS patterned NF substrate, demonstrated ~130 mV (40 %) of lower $\eta_{10}$ overpotentials in the HER over the same loading of electrocatalysts on the pristine NF electrode along with significantly improved stability. Similarly, RuO$_2$ electrocatalyst powder, the model OER electrocatalyst, loaded on the LIPSS patterned NF substrate, required ~100mV (25 %) lower $\eta_{10}$ overpotential in the OER with significantly enhanced stability. Importantly, when two LIPSS patterned electrodes were assembled simultaneously as anode and cathode to make a cell, it requires 330 mV of lower electric potential over a similar cell made of pristine NF electrodes to drive 10 mA/cm$^2$ in overall water. The capability of the LIPSS patterned electrodes to operate at significantly reduced electric potentials is demonstrated in a range of electrolytes including alkaline, acidic, neutral, and seawater. The present work demonstrates a single step, fast, and physical approach of electrode surface patterning to boost electrocatalytic fuel generation performance and can be applied to any metal and semiconductor catalysts for reducing the required electrical power in a range of electrochemical reactions.

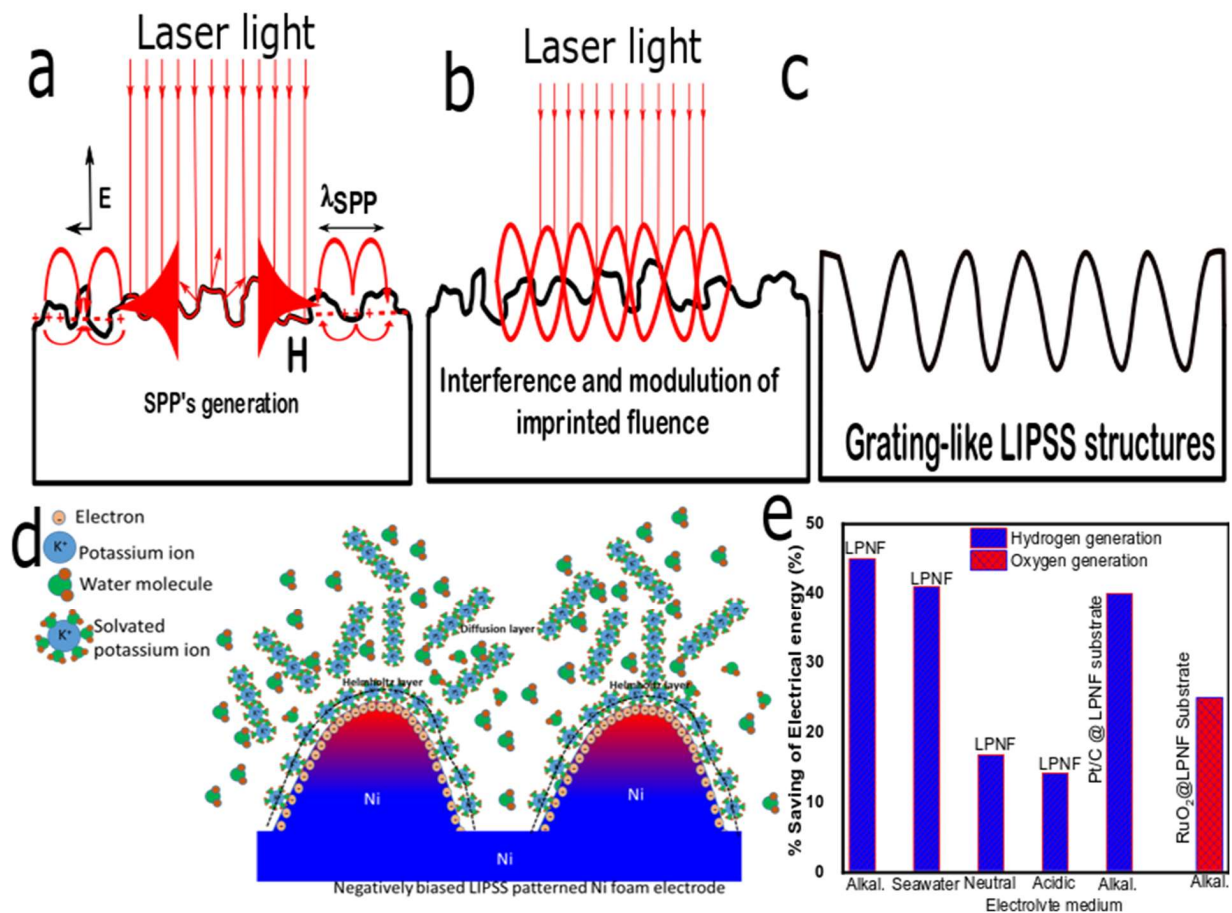

**Fig. 1**: **Schematic illustrations of LIPSS formation, localized electric field induced enhancement in reagent concentration and experimental results in decrease in electrical energy:** Schematic of LIPSS formation (a) surface plasmon polariton (SPP) generation and propagation parallel to the surface, (b) interference between the incident light wave and SPP wave to modulate the light intensity, and (c) imprinting of the interference pattern thorough selective ablation of material. (d) Schematic of the Gouy-Chapman-Stern electrical double layer model to represent mechanism behind localized electric field induced enhancement in the reagent concentration, and (e) experimental results showing percentage decrease in the require electrical potential to generate 10 mA/cm$^2$ of current density using the LIPSS patterned electrodes in different electrolytes.

## 2. Experimental section

### 2.1 COMSOL Multiphysics simulations

To calculate the electric field generated by our LIPSS patterned on Ni foam electrodes, we used Finite Element based solver COMSOL Multiphysics. Electrostatics module was used to compute the electric (E = −∇V) under a specific electrode bias potential. Two-dimensional geometry was built to represent the nanorod and nanoparticle structures used in this work. Triangular meshes were used for all simulations. Meshes were set to be the densest at the surface of the electrodes.

## 2.2 Materials

Ni foam (thickness 1.5mm, porosity ~98%) was used as received. Deionized (DI) water (resistivity: 18.3MΩ) was used, as the medium for the preparation of an aqueous electrolyte solution.

## 2.3 Fabrication on micro/nanostructures on Ni-foam

The fabrication of Laser-Induced Periodic Surface Structures (LIPSS) on Ni foam was performed with the laser at 800 nm, 40 fs pulses from Ti: Sapphire laser (Spitfire Ace, Spectra-Physics) at 1 kHz repetition rate. The maximum pulse energy output from the laser system is 6 mJ, which was attenuated using the combination of a half-wave plate and a linear polarizer. The fast fabrication of LIPSS on the surface of 0.5 cm x 0.5 cm nickel foam samples could do with the cylindrical lens of 50 mm focal length and an XYZ translational stage. Samples were scanned across the optical axis by fixing at the computer-controlled translational stage. Provided, the laser beam was irradiated on the surface at normal incidence. The fabrication was carried out with 95uJ, 236uJ, 460uJ, 600 µJ, and 700 µJ pulse energy and 0.5 mm/s scanning speed at the focal point. For convenience, the so-fabricated were Laser-Induced Periodic Surface Structures (LIPSS) on Ni foam abridged as LPNF-95, LPNF-236, LPNF-460, LPNF-600, and LPNF-700.

## 2.4 Synthesis of Pt/C and RuO$_2$ electrodes @NF and LPNF

To prepare the electrocatalyst inks, commercial Pt/C 20 % or RuO$_2$ (5 mg) catalysts powder was dispersed in 1 mL of ethanol and DI water (V=1:1) mixture. The ink of each catalyst was prepared by first ultrasonication the solution for 1h, followed by the addition of 36 µL Nafion solution (5 wt%) into the mixture as adhesive and ultrasonication of the mixture again for 30 min to form a homogeneous suspension. Then the prepared slurry was painted onto the BNF and LPNF-600 substrates and finally dried at room temperature. For our simplicity, the as-synthesized electrocatalysts abbreviated as Pt@BNF, Pt@LPNF-600, RuO$_2$@BNF, and RuO$_2$@LPNF-600.

## 2.5 Characterization of electrocatalysts

The x-ray diffraction (XRD) measurements of different samples were carried out using Bruker D8 Focus X-ray diffractometer with Cu-Kα (λ = 1.5406 Å) radiation operating at a voltage of 40 kV and a current of 30 mA. The scanning electron microscopic (SEM) images of as prepared electrocatalysts were measured using a HITACHI S4800 scanning electron microscope. The X-ray photoelectron spectra were recorded using Thermo Escalab 250XI X-ray photoelectron spectroscopy with AlKα X-ray source.

## 2.6 Electrochemical Measurements

The electrochemical measurements were performed on a BioLogic VMP3 multichannel workstation with a three-electrode system, where a Pt wire, structured electrocatalysts (0.5 cm×0.5 cm), and a saturated calomel electrode (SCE) were used as a counter, working and reference electrodes, respectively. An aqueous solution of 1M KOH was used as an electrolyte for bare and all laser patterned samples, the electrochemical measurements. Each of the working electrodes was pre-scanned for 50 Cycles of cyclic voltammetry (CV) with the scan rate 40mVs$^{-1}$ before performing linear sweep voltammetry (LSV) curves. The LSV curves were measured by sweeping voltage in the range of -0.2 to -1.6 V vs SCE electrode with the scan rate of 10 mVs$^{-1}$. The expression $E_{RHE} = E_{SCE} + E^0_{SCE} + 0.0592 * pH$, where $E^0_{SCE} = 0.242\ V$, was used to translate V *versus* SCE to V *versus* reverse hydrogen electrode (RHE). Electrochemical impedance spectroscopy (EIS) was measured at a dc overpotential of 0.341V vs RHE (For HER) superimposing a small alternating voltage of 10 mV over the frequency range of 10 mHz to 1 MHz. The CV curves were further measured in the non-Faradaic region of potential from 0.84V to 1.04V (*versus* RHE) for HER with different scan rates (from 20 mVs$^{-1}$ to 120 mVs$^{-1}$ for HER, to estimate the double-layer capacitance ($C_{dl}$) and electrochemically active surface area (ECSA). The slope of the difference in the cathodic and anodic current densities ($\Delta J = J_c - J_a$) with the scan rate resulted in $C_{dl}$. The ECSA values were then calculated using the expression $E\ CSA = C_{dl(catalysts)}/C_{dl\_NF}$; where $C_{dl-NF}$ is the double layer capacitance for the bare Ni foam. Here, we used the $C_{dl}$ value of the bare Ni foam instead of the general specific capacitance ($C_s$) to exclude the effect of the larger capacitance value of bare NF. All the electrochemical measurements were performed at room temperature. The long-term durability measurements were done using 24 hours of chronopotentiometry (CP) and LSV curves before and after the CP measurements.

## 3. Results and Discussion

Schematic illustration and experimental setup for the femtosecond laser writing of LIPSSs on a Ni foam electrode are shown in **Fig. 1**(a-c) and **Fig.** S1 respectively. In a typical LIPSS experiment, a burst of linearly polarized femtosecond laser pulses is focused on the target surface. Depending on the instantaneous surface roughness, a fraction of absorbed radiation excites surface plasmon polaritons (SPPs) wave on the sample surface (**Fig. 1**(a)), while the rest of the absorbed energy gets transferred to the lattice through electron-phonon interaction. According to electromagnetic theory, an interference between the incident radiation with the SPPs results in spatial modulation of the intensity at the target surface[37,38] (**Fig. 1**(b)). Finally, the interference pattern gets imprinted through selective ablation of the material to form LIPSSs **(Fig. 1**(c)). Formation of LIPSSs is a sequential phenomenon where the structures formed via a preceding laser pulse guides the structures to be formed with the next laser pulse. Therefore, for given pulse width, the geometric parameter of a LIPSS pattern depends on the number of pulses at a given place and the laser pulse energy. We fabricated five different LIPSSs patterned nickel foam (LPNF) electrodes using 95, 236, 460, 600, and 700 µJ/pulse laser energy and 1 mm/s scanning speed. Hereafter, the LIPSSs patterned electrocatalysts are referred as LPNF-95, LPNF-236, LPNF-460, LPNF-600, and LPNF-700, respectively while the bare NF electrode is presented as BNF. When a LIPSSs patterned electrode will be placed in an alkaline medium it would form an electrostatic arrangement at the electrode-electrolyte interface and in the electrolyte. A schematic of the Gouy-Chapman-Stern electrical double layer with Helmholtz and diffusion layers is shown in **Fig. 1**(d) and **Fig.** S2.

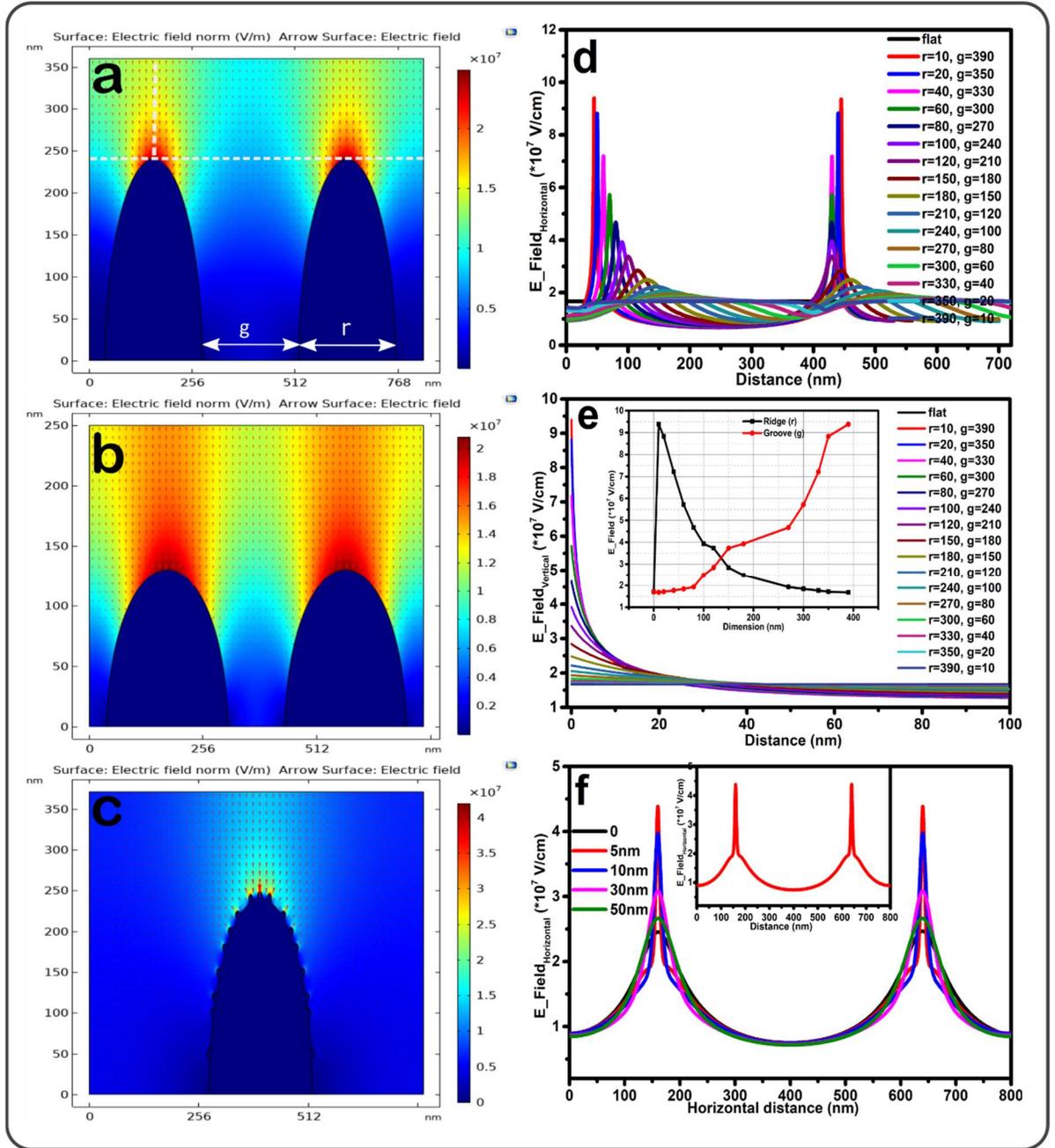

**Fig. 2: Finite Element Model (FEM) simulation for the spatial distribution of the electric fields around the ridges and grooves of -1 V biased LIPSS patterns;** Electric field mapping without considering nanoparticles on the ridges and grooves with geometric parameters (a) g:180 nm, r:150 nm, and h: 240 nm and (b) g:100 nm, r:240 nm and h: 125 nm, and (c) with the addition of 5 nm nanoparticles at the ridge, (d) Electric field intensity along the (d) vertical and (e) horizontal dotted lines, shown in (a), for LIPSSs patterns of different geometric parameters without nanoparticles, and (f) Spatial variation in the electric field intensity along the horizontal dotted line for the LIPSS pattern shown in (a) with the addition of spherical nanoparticles of different sizes.

The Helmholtz layer consists of a monolayer of surface adsorbed solvated cations on the nanostructured electrode surface while the diffusion layer consists of cations and anions that diffuse freely in the electrolyte and forms a concentration gradient. [36] Localization of electric field near the curved surfaces of a biased electrode forms hotspots. These hotspots can pull water molecules, adsorbed on the cations' surfaces (**Fig. 1**(d)) and **Fig.** S2), from the bulk of the electrolyte to their surfaces to dissociate them at much lower applied electric potential to produce $H^+$ and $OH^-$ reagents for hydrogen and oxygen generations, respectively. The intensity of the hotspots depends on the geometric parameters of the LIPSSs pattern and the size and density of NPs present at the surface. **Fig. 1**(e) shows a histogram summarizing results of percentage decrease in the required electrical energy to drive 10 mA/cm$^2$ current density in different electrolytes for hydrogen (blue bars) and oxygen (red bars) generations.

First, we used the FEM simulation to explore the interconnection between the surface topography (dimensions of ridges and grooves) and morphology (size and density of NPs) of the LIPSSs patterns and intensity of the hotspots. A semi-elliptical periodic structure with a ridge width r and a groove width g is used to represent the cross-section of a LIPSS patterned electrode. The electric field mappings around negatively biased LIPSSs patterned electrodes with geometric parameters; g:180 nm, r:150 nm, and h: 240 nm (**Fig. 2**(a)) and g:100 nm, r:240 nm, and h: 125 nm (**Fig. 2**(b)) are shown in **Fig. 2**. Because of the electrostatic repulsion, the free electrons in the negatively biased electrode get migrated to the region of high curvature (near the tip) resulting in the generation of a highly localized electrostatic field. The localized electric field intensity for the LIPSSs patterns with the radii of ridge curvatures 120 nm (**Fig. 2**(a)) and 135 nm (**Fig. 2**(b)) are ~2.5×10$^7$ and 2.0×10$^7$ V/cm respectively. The spatial variations in the horizontal and vertical components of the electric fields for different LIPSSs' parameters are demonstrated in **Fig. 2**(d) and **2**(e), respectively. Compared to the flat surface (E-field;1.7×10$^7$), the LIPSS patterned electrode with r= 10 nm (E-field;9.5×10$^7$) can generate ~ 5.58 times intense hotspots. Moreover, NPs on the surface of ridges and grooves can further enhance the intensity of hotspots (**Fig. 2**(c)). To understand the effects of the size of NPs and their density on the E-field localization, we performed extensive simulation (See supporting information **Fig.** S3 – **Fig.** S6). We chose g:180 nm, r:150 nm, and h:240 nm periodic structure of Ni, shown in **Fig. 2**(b), and put spherical Ni NPs on the ridges. The presence of Ni NPs of 5 nm diameter on the ridges shows enhancement in the E-field intensity from 2.5×10$^7$ to 4.5×10$^7$ V/cm (**Fig. 2**(a) *versus* **Fig. 2**(c); see **Fig.** S4 (a) **and**

S4(b)). **Fig. 2**(f) shows the spatial distribution of the horizontal component of the E-field for the bare LIPSSs (shown in **Fig. 2**(a)) and the same LIPSSs decorated with Ni NPs of different diameters. The presence of NPs of 5 and 10 nm diameters localizes the E-field strongly as compared to NPs of larger diameters. For example, the NPs of 50 nm diameter have a negligible effect in the enhancement of E-field intensity.

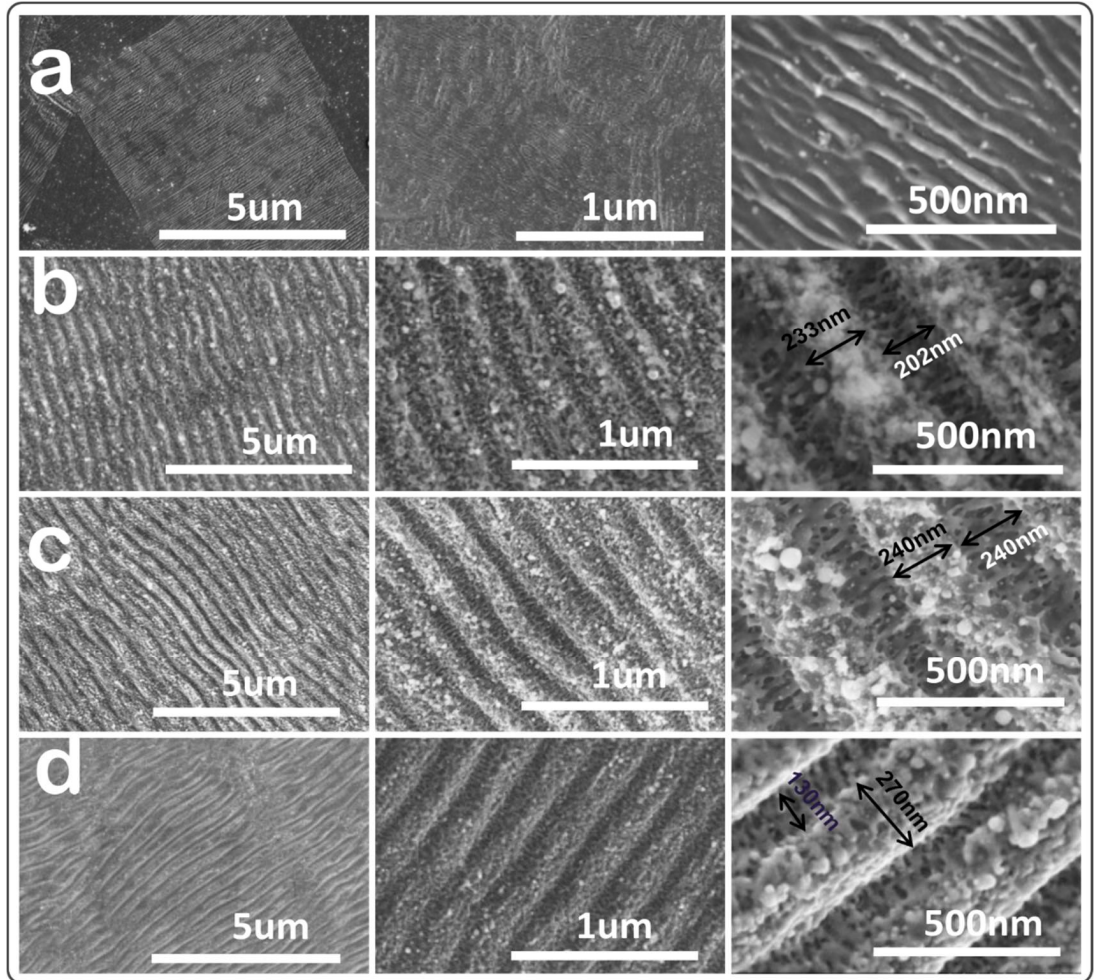

**Fig. 3**: **The surface morphology and topography of LIPSS patterned electrodes:** The SEM images of LIPSS patterned Ni foam (LPNF) electrode using (a) 236, (b) 460, (c) 600, and (d)700 µJ/pulse of femtosecond laser pulse energy having 1kHz repetition rate. The focused laser beam was scanned on the Ni foam target with the scanning speed

The scanning electron microscopy (SEM) images (**Fig. 3**) of the LIPSSs patterned NF electrodes can be used to extract the average dimensions of ridges and grooves and the size of the NPs on their surfaces. At a low laser pulse energy (236 µJ/pulse), the LIPSSs pattern is irregular and short length (**Fig. 3**(a) middle panel) with 60 and 100 nm average dimensions of ridges and grooves, respectively, with very low density of NPs. With an increase in the laser pulse energy, up to 600 µJ/pulse, the ratio of groove to ridge (g/r) widths increases (see supporting information Table S1) with an increase in the regularity of the LIPSSs pattern and lengths of the ridges and grooves (See **Fig. 3**(b) and **Fig. 3**(c); left panels). For example; the length and regularity of the LIPSSs pattern at LPNF-460 (Fig. 2(b); left panel) is certainly higher as compared to the LIPSSs pattern at the LPNF-236 electrode (**Fig. 3**(a); left panel), but lower than the pattern formed at the LPNF-600 electrode (**Fig. 3**(c); left panel). From these images, we can see that the depth of the grooves and density of NPs on the surface of the LIPSSs are larger at higher laser pulse energies (**see Table** S1). The LIPSSs pattern at the LPNF-600 electrode has 240 nm average sizes of grooves and ridges where the surface is covered with spherical NPs of 16.6±8.4 nm average diameter (see supporting information **Fig.** S7). Further increase in the laser pulse energy to 700 µJ/pulse possibly results in the melting of smaller sized NPs and solidification of the melted material on the sides of ridges, consequently formation of wider rides (270 nm), narrower (130 nm) grooves, and larger size of NPs at the surface. The surface of the LPNF-700 electrode has hemispherical NPs with an average diameter of 39.2 ±3.2 nm (see supporting information **Fig.** S7). The X-ray diffraction spectra of the LPNF-600 and BNF electrodes show that the surface chemistry of the electrodes is retained after laser surface treatment (see supporting information **Fig.** S8).

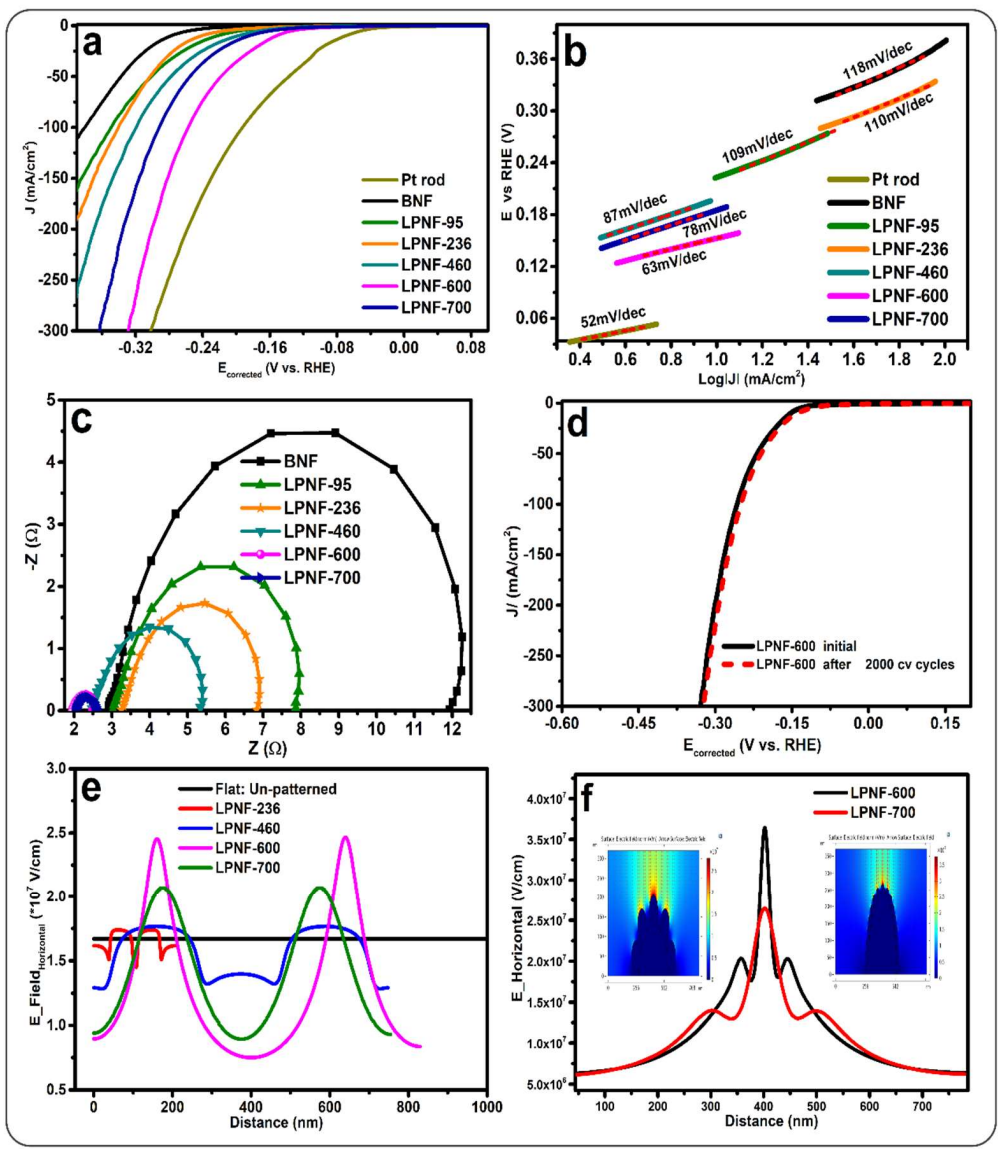

**Fig. 4. Demonstration of LEFIRC effect induced reduction in the overpotential for electrochemical hydrogen generation when the LIPSS patterned electrode is directly used as electrocatalyst:** HER performance of on BNF and different LPNF electrodes with Pt/C electrode as a reference in 1 MKOH solution at room temperature: (a) the LSV curves and corresponding (b) Tafel slopes and (c) Nyquist plots, and (d) the LSV curves before (solid black) and after (dotted red) 2000 CV cycles, (e) the FEM simulation for the distribution of electric field intensity along the horizontal line for the different LIPSS patterns without considering the effects of nanoparticles, and (f) comparison of electric field localization at the ridges of LPNF-600 and LPNF-700 electrodes with the addition of NPs of average sizes 15nm, and 40nm respectively. Inset (corresponding 2D mapping of electric field localization).

To verify the proposed LEFIRC effect induced enhancement in the electrochemical fuel generation performance, we first tested the HER activity of the as-fabricated LIPSSs patterned electrodes in an alkaline (1M KOH) medium and later in other electrolytes. **Fig. 4**(a) shows the linear sweep voltammetry (LSV) curves, with iR compensation, for different LIPSSs patterned NF and BNF electrodes along with the LSV curve of a Pt-rod electrode for reference (*see* non-iR compensated LSV curves in supplementary information **Fig.** S10). At a given applied potential, the current density increases with an increase in the groove to the ridge ratio (g/r) of the LIPSSs patterned electrode and has a maximum current density value for the LPNF-600 electrocatalyst having the highest g/r value (see supporting information **Table** S2 **and Table** S3). The LPNF-700 electrocatalyst shows comparatively lower current density due to a decrease in the electric field intensity from wider ridges, narrower grooves, and larger sizes of NPs on the LIPSSs surface (see supporting information **Fig.** S7 and **Table** S1). From the LSV curves, we can see that LPNF-600 is the best HER performer with the highest current density at a given overpotential. The origin of enhanced current density, at a given potential, can be a combination of increased electrochemical surface area (ECSA) and the LEFIRC effect. The LIPSS patterning can alter the ECSA value of the electrode thus corresponding current density. To evaluate the contribution of the LEFIRC effect and ECSA on the enhanced current density, we calculated ECSA values of the LIPSSs patterned electrodes using the corresponding double-layer capacitance ($C_{dl}$) values extracted from corresponding cyclic voltammetry (CV) curves recorded at different scan rates in non-Faradaic regions (see supporting information **Fig. S11-Fig.** S13 and **Table** S4). The ECSA value for the LPNF-95 electrocatalysts sample (0.81) is relatively lower than the BNF (1.0) electrocatalyst indicating that the fs laser pulses remove pre-existing surface textures at the NF and made it relatively smoother with ultrafine periodic patterns (see supporting information **Fig.** S15). In contrast, at higher laser pulse energies, the ECSA value of the LIPSSs patterned electrodes increases with an increase in the laser power resulting in 1.44 and 2.19 values for the LPNF-600 and LPNF-700 electrodes, respectively.

To explore the LEFIRC's contribution in the electrocatalytic performance of the LIPSSs patterned electrodes, we normalized the LSV curves with corresponding ECSA values (see supporting information **Fig.** S14 (c, d)). Comparing the $\eta_{10}$ overpotential values between the original (**Fig. S14(b)**) and corresponding ECSA normalized LSV curves (**Fig. S14(d)**), we estimated that the ECSA contribution is almost negligible in the LPNF-95, LPNF-236, and LPNF-460 LIPSSs patterned electrodes and have about 3 % and 12 % contributions in the LPNF-600 and LPNF-700 electrodes, respectively. From the detailed investigations, we can claim that the LEFIRC effect is the main cause of higher observed electrocatalytic hydrogen generation performance in all the LIPSSs patterned electrodes. The LEFIRC effect at the LIPSS patterned electrodes reduces $\eta_{10}$ overpotential values in the range of 13 to ~43 %. The LIPSSs patterns at the LPNF-600 electrode reduces the required electrical energy by ~ 45.4 % to achieve a hydrogen generation rate of ~$3\times10^{16}$ molecules cm$^{-2}$s$^{-1}$ (current density: 10 mA/cm$^2$) via the synergistic effects of LEFIRC and ECSA enhancements where ~ 42.4 % contribution is purely from the LEFIRC effect.

As shown in **Fig. 4**(b), the Tafel slopes for the BNF, LPNF-95, LPNF-236, LPNF-460, LPNF-600, and LPNF-700 electrocatalyst electrodes are 118, 109, 110, 87, 63, and 78 mV/dec, respectively (see supporting information Table S2 and Table S3). The Tafel slope for the most efficient LIPSS patterned electrode, LPNF-600, is 46.6 % lower as compared to the BNF. The electrochemical impedance spectroscopy (EIS) (**Fig. 4**(c)) of the electrocatalyst samples shows that the LPNF-600 (2.5 Ω) electrocatalyst has significantly lower charge-transfer resistance as compared to the LPNF-700 (2.6Ω), LPNF-460 (5.3 Ω), LPNF-236 (6.8 Ω), LPNF-95 (7.8 Ω), and the bare NF (11.4 Ω) electrocatalysts resulting in a higher rate of electron transfer from the catalyst to reduce the adsorbed H* in the Volmer step of the HER. The localization of the electric field of the order of ~$4.0\times10^7$ V/cm can increase the concentration of H* radicals on the catalyst surface through the electrostatic attraction of the solvated potassium ions and dissociation of water molecules adsorbed on their surface (**Fig. 1**(f) and **Fig. S2**). Moreover, the long-term durability of an electrocatalyst is an important parameter to evaluate its performance. A complete overlap of the LSV curve recorded after 2000 CV cycles **(Fig. 4**(d)) with the initial one demonstrates excellent stability of the LPNF-600 electrocatalyst cathode in an alkaline medium. The LSV curves before and after 2000 CV cycles for the LPNF-460 and LPNF-700 electrocatalyst samples demonstrate their excellent stability (Fig. S16) in the alkaline medium. To understand the reason behind the higher electrochemical performance of the LPNF-600 electrocatalyst over others, we extracted geometric parameters of the LIPSS patterns for each sample and measured the size of nanoparticles at ridges and grooves from corresponding SEM images. These data are used to calculate the electric field profiles (see supporting information **Fig. S9**) for each sample (**Fig. 4**(e)). The trend of the calculated electric field intensity supports the observed trend in the electrochemical performance. Moreover, the smaller size of nanoparticles (av: 16.6±8.4 nm) at the surface of LPNF-600 electrocatalyst better localize ($3.75\times10^7$ V/cm) electric field, thus has a higher LEFIRC effect induced electrochemical fuel generation performance over the LPNF-700 electrocatalyst covered with 39±3.2 nm average size of particle generating $2.7\times10^7$ V/cm of the electric field (**Fig. 4**(f)).

According to the Gouy-Chapman-Stern model, (Schematic **Fig. 1**(f) and **Fig. S2**) the density of water molecules that can be electrostatically transported to the electrode surface through solvation depends on the size, charge, and mobility of the carrier cation/anion of the electrolyte. For example, a cation/anion with higher valency experiences a higher electrostatic force, but the rate of transport of these ions to the electrode depends on their mobility. To explore the roles of electrolytes on the LEFIRC effect induced reduction in the required electrical power in fuel generation, we tested the HER performance of the LPNF-600 electrode in seawater ($4\times 10^4$ ppm NaCl), acidic (0.5 M $H_2SO_4$), and neutral (deionized water 18 MΩ-cm) electrolyte media (see supporting information **Fig.** S17 and S18). The η₁₀ overpotentials for the LPNF-600 electrode

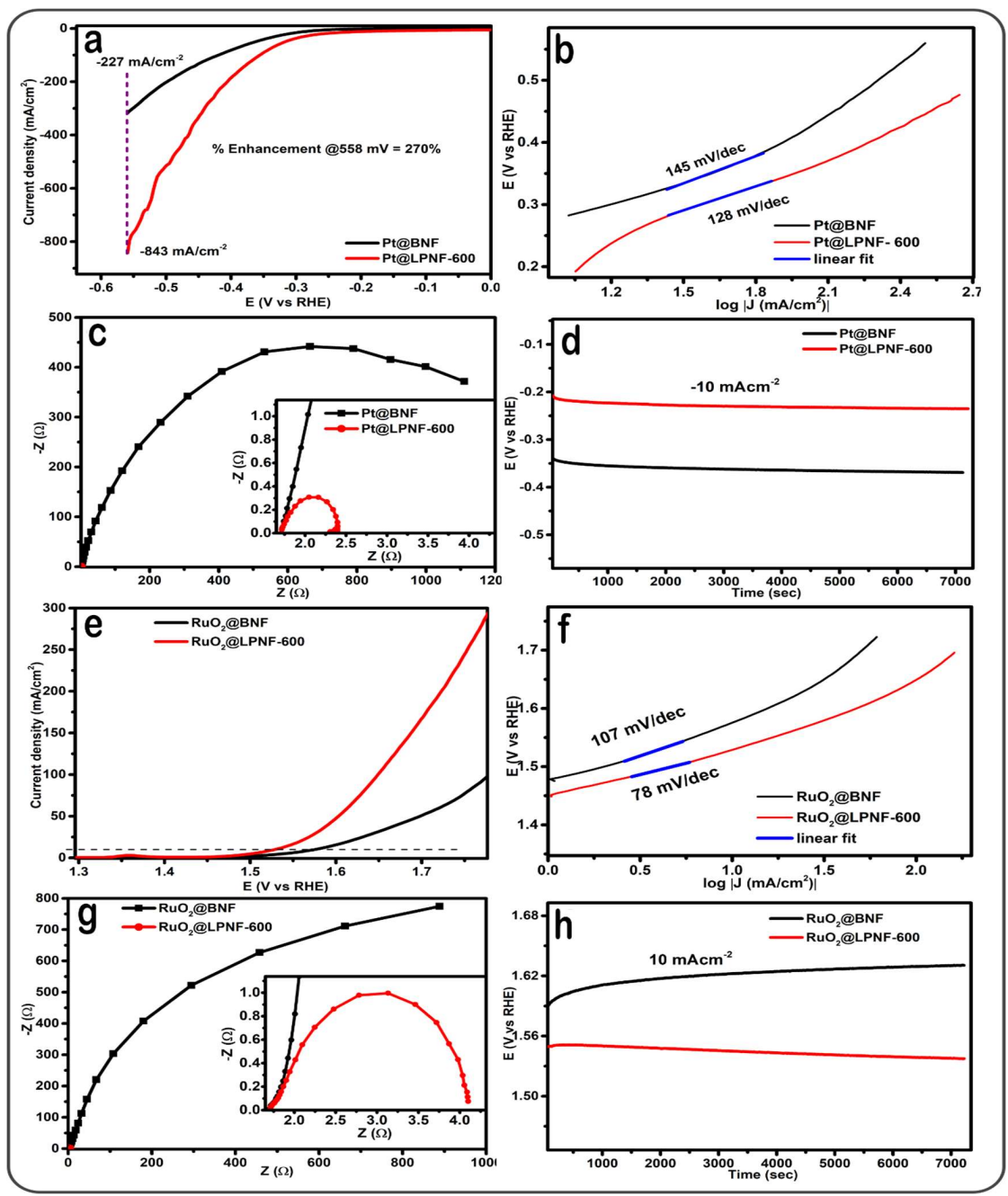

**Fig. 5: Demonstration of LEFIRC effect induced reduction in the electrochemical fuel generation potential when the LPNF-600 sample is used as cathode or anode substrate to support Pt/C or RuO₂ electrocatalyst:** (a-d) HER and (e-h) OER measurements for Pt/C and RuO₂ inks loaded on BNF and LPNF-600 substrates in 1 MKOH solution at room tempereture.

in seawater, acidic, and neutral electrolytes are 335, 326, and 748 mVs those are ~ 41, 14, and 17 %

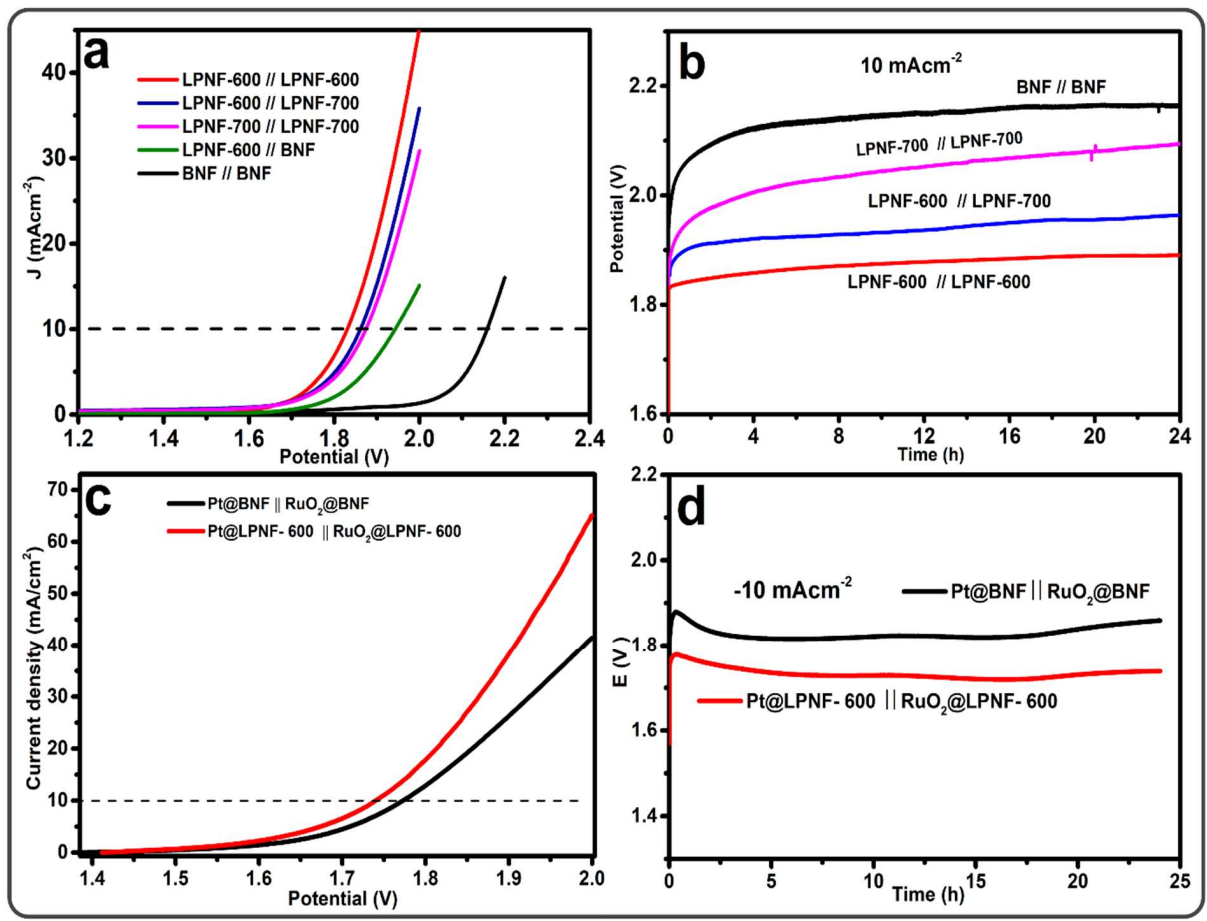

**Fig. 6: Demonstration of LEFIRC effect induced reduction in the overall water splitting potential when the LIPSS patterned electrodes are used as anode and cathode: (a & b)** The LPNF surfaces are directly used as electrocatalysts for anode and cathode (a) the LSV curves for overall water splitting from different combinations of LPNF and BNF electrodes, and (b) corresponding chronopotentiometry curves for 24 hours. (c & d) A pair of LPNF-600 surfaces are used as substrates to support Pt/C and $RuO_2$ electrocatalysts to make cathode and anode, respectively of the cell along with a similar cell made with the BNF substrates for comparison (c) the LSV curves for overall water splitting and (d) corresponding chronopotentiometry curves for 24 hours.

lower as compared to the corresponding overpotentials of the BNF electrode.

Periodically patterned substrates, known as metasurfaces, are widely used in plasmonics and nanophotonics to boost performances of photodetectors, solar cells, and plasmonic sensors as well as in guiding and manipulating the characteristics of light [39–48]. However, unfortunately periodically patterned surfaces are never used as substrates to boost the performance of electrocatalyst powders. To extend the application of the LIPSSs patterned electrode as a substrate that not only can support a powder electrocatalyst or ink but can also enhance its performance via the LEFIRC effect, we selected the LPNF-600 electrode and loaded it with Pt/C or $RuO_2$ (**Fig. 5**) powder electrocatalyst to enhance their HER or OER performances, respectively in an alkaline solution. The LSV curves for the BNF and LPNF-600 electrodes, loaded with 0.25 mg/cm$^2$ of Pt/C (ratio; 2:8) electrocatalysts powder, show 302 and 180 mVs of $\eta_{10}$ overpotentials, respectively resulting in ~40.4 % decrease in the required electrical power (**Fig. 5**(a) and **Fig.** S19) to drive 10 mA/cm$^2$ of current density. At an applied electric potential of 558 mV, the hydrogen generation performance of the LPNF-600 substrate is enhanced by 270 % over the BNF substrate. The corresponding Tafel slope for the LPNF-600 substrate (128 mV/dec) is 11.7 % smaller compared to the BNF substrate (145 mV/dec) (**Fig. 5**(b)). The enhanced performance of the LPNF-600 substrate is also evidenced by a decrease in its charge transfer resistance to a small value of 2.25 Ω as compared to > 1kΩ resistance for the BNF substrate (**Fig. 5**(c)). The chronopotentiometry curves, at 10 mA/cm$^2$ current density, for the LPNF-600 and BNF substrates (**Fig. 5**(d)) further show long-term stability of the LIPSS patterned substrate in the E-filed localization.

In the previous sections, we presented the capability of the LIPSS patterned electrodes as cathodes to demonstrate the LEFIRC effect-induced reduction in the electrochemical fuel generation potential. Similarly, we present the performance of the LIPSSs patterned NF substrate, as an anode, to support and boost the activity of an OER electrocatalyst in an alkaline electrolyte (**Fig. 5** (e-h)). The LSV curves for the BNF and LPNF-600 substrates, loaded with 0.25 mg/cm$^2$ of $RuO_2$ electrocatalysts power, demonstrate that the LEFIRC effect reduces the $\eta_{10}$ overpotentials from 400 mV (BNF) to 300 mV (LPNF-600) resulting in a 25 % reduction in the required electrical power (**Fig. 5**(e) and **Fig.** S20 (b)) to drive 10 mA/cm$^2$ of current density. At an overpotential of 750 mV, the oxygen generation performance of the LIPSS patterned substrate is enhanced by 211 % over the bare substrate (**Fig.** S20(a)). The Tafel slope for the LPNF-600 substrate (78 mV/dec) is 27.1 % lower as compared to the BNF (107 mV/dec) substrate (**Fig. 5**(f)). The enhanced performance of the LPNF-600 substrates is also evidenced by a decrease in its charge transfer resistance to a small value of ~4Ω as compared to > 1kΩ resistance for the bare NF electrode (**Fig. 5**(g)). Moreover, the chronopotentiometry curves at 10 mA/cm$^2$ current density for the LPNF-600 and BNF substrates **(Fig. 5**(h) further demonstrates the long-term stability of the LIPSS pattern substrate in the localization of E-field and oxidation of $OH^-$ ions for oxygen

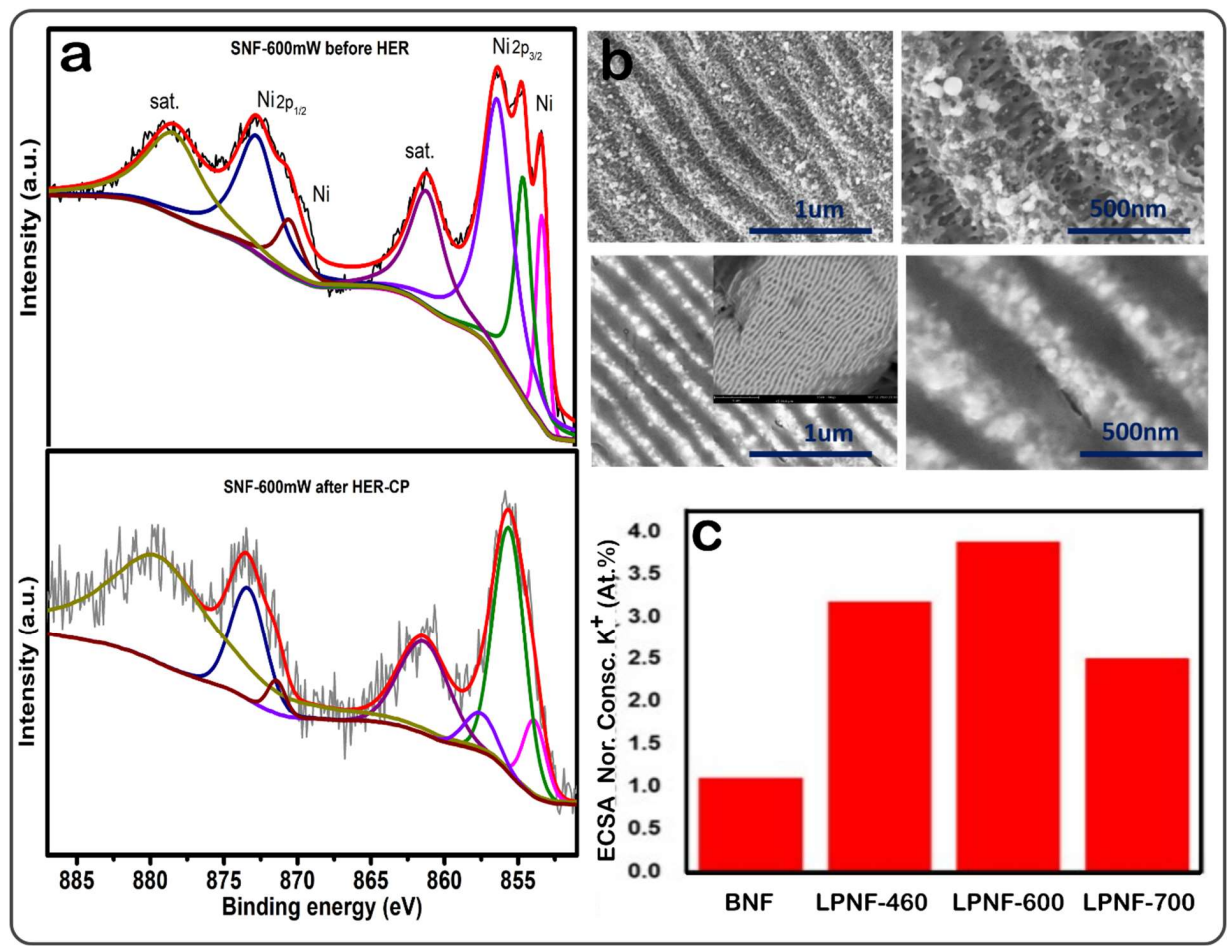

**Fig. 7:** (a) XPS spectra, (b) SEM images, and (c) ESCA normalized atomic concentration of $K^+$ ions at the surface of – 1 V (vs RHE) biased LPNF electrodes in 1M solution of KOH.

generation.

Furthermore, we constructed two-electrode electrochemical cells (ECCs) using different combinations of cathode and anode materials for overall water splitting **(Fig. 6**(a)). The electrochemical cell LPNF-600 || BNF (1.94 V) requires about 210 mV smaller electric potential to drive 10 mA/cm$^2$ of current density as compared to the BNF || BNF cell (2.15 V). Similarly, the LPNF-600 || LPNF-600 (1.82 V), LPNF-600 || LPNF-700 (1.86 V), and LPNF -700 || LPNF-700 (1.88 V) electrochemical cells require 15.3 %, 13.5 %, and 12.6 % lower electric potential to generate 10mA/cm$^2$ of current density as compared to the BNF || BNF cell (2.15 V) **(Fig. 6(a)).** The chronopotentiometry curves (**Fig. 6**(b)) of the ECCs made of a pair of the LIPSS patterned

electrodes show that these cells can operate at significantly lower (from 160 to 270 mV) potentials for 24 hours, thus maintaining the surface patterns, morphology, and chemical composition. A two-electrode cell (Pt/@LPNF-600 || RuO$_2$@LPNF-600) made with a pair of LPNF-600 substrates, where the first one loaded with 0.25 mg/cm$^2$ of Pt/C (ratio; 2:8) powder serves as a cathode for the HER and another loaded with 0.25 mg/cm$^2$ of RuO$_2$ powder acts as an anode for the OER is tested for overall water splitting. For comparison, a similar cell (Pt/@BNF || RuO$_2$@BNF) made with the loading of the same amount of electrocatalyst powders on the BNF substrates is also tested. It is evident from **Fig. 6**(c) that the Pt/@LPNF-600 || RuO$_2$@LPNF-600 cell operates at significantly lower potential (1.74 V) as compared to the Pt@BNF || RuO$_2$@BNF cell (1.83 V) to drive 10mA/cm$^2$ of current density. The chronopotentiometry measurements of these two cells demonstrate the durability of the LIPSS patterned substrates and NPs on their surfaces against overall water splitting in an alkaline solution.

Moreover, structural, compositional, and morphological integrities of an electrocatalyst are important parameters to ensure its long-term durability. For this purpose, we performed XPS, SEM, EDAX, and XRD measurements of the LPNF-600 sample before and after 24 hours of HER test (**Fig. 7**, supporting information Fig. S8 and Fig. S20-S28). The Ni$^0$ peaks at 852.7 and 869.9 eVs, present in the initial LPNF-600 sample (top panel of **Fig. 7**(a)), disappeared after 24 hours of the HER test (bottom panel of **Fig. 7**(a)) showing that surface Ni atoms oxidized to form NiO, an active component for HER and OER enhancements. The high-resolution O1s spectrum (see supporting information Fig. S 22) further verifies conversion of Ni to NiO after the HER process. The SEM investigations before (**Fig. 7**(b) top panels) and after (**Fig. 7**(b) bottom panels) show integrity of the majority of the LIPSSs and NPs on the ridges (see supporting information **Fig. S23-S26**). However, we noticed the narrowing of the ridges and widening of the grooves in all the LPNF samples after the HER tests. This structural change increases the g/r ratio, thus positively affects the electrochemical performance through the LEFIRC effect. For example, the potential required to drive 10mA/cm$^2$ of current density in LPNF-600|| NF cell decreased by 140 mV after 24 hours of CP test. In contrast to the BNF (see supporting information **Fig.** S27(a, c)) substrate, the performance of the HER and OER performances of the LPNF-600 substrates significantly increase after 24 hours of CP tests (see supporting information Fig. S27-S28). To support our hypothesis of LEFIRC effect induced enhancement in the electrochemical fuel generation performance, we tested the concentration of K$^+$ ions adsorbed at the surface of -1 V (versus RHE) biased LPNF electrodes (see Supporting Information **Fig.** S28) in 1M aqueous medium of KOH using EDAX spectroscopy. The surface density of adsorbed ions at the electrode surface should be proportional to the intensity of the localized electric field. As can be seen from **Fig. 7**(c), the surface density of adsorbed K$^+$ ions at the LPNF electrodes is proportional to the ECSA normalized current density in the HER LSV curve (**Fig.** S14).

## 4. Conclusion

In summary, we fabricate LIPSSs patterns on Ni foam (LPNF), one of the widely used electrode materials, and optimize geometric parameters for localized electric field-induced modulation in reagent concentration (LEFIRC) at the electrode surface. The patterned electrodes have hierarchal structures with periodic ridges and grooves of the order of 100 to 300 nm in widths covered with nanoparticles (NPs) of 3 to 94 nm diameters. The optimized LPNF electrode demonstrated ~45.3 % reduction in the required electrical power to achieve a hydrogen generation rate of ~3×10$^{16}$ molecules cm$^{-2}$s$^{-1}$ (current density: 10 mA/cm$^2$) via the synergistic effect of LEFIRC and ECSA enhancements where ~ 42.4 % contribution is purely from the LEFIRC effect. The capability of the LPNF electrodes to operate at significantly reduced electric potentials is also demonstrated in a range of electrolytes including acidic, alkaline, neutral, and seawater. The $\eta_{10}$ overpotentials for the LPNF electrodes in seawater, acidic, and neutral electrolytes are 335, 326, and 748 mVs those are ~ 41, 14.2, and 16.8 % lower as compared to the corresponding overpotentials of the bare NF (BNF) electrodes. Furthermore, the applicability of the LPNF electrodes as substrates to support powder electrocatalysts and boost their fuel generation performance is also demonstrated. Pt/C powder loaded on the LPNF substrate demonstrated a 40.4 % reduction in the required electrical potential to drive 10 mA/cm$^2$ of current density with 270 % higher current density at 558 mV potential over a BNF substrate with the same amount of catalyst loading. Similarly, RuO$_2$ powder loaded on the LPNF substrate demonstrated a 25 % reduction in the $\eta_{10}$ overpotentials and 211 % enhancement in the current density at 1.98 V applied potential. Importantly, when two LIPSS patterned (LP) electrodes were assembled simultaneously as anode and cathode, it requires ~ 330 mV of lower electric potential over a similar cell made of BNF electrodes to drive 10 mA/cm$^2$.

The demonstrated operation of the LPNF electrocatalysts at significantly lower electrical energies makes it clear that the femtosecond laser patterning approach has the potential to deliver next-generation smart catalysts. While we tested the LIPSS patterning of Ni foam electrodes, other materials can be LIPSSs patterned to enhance their electrochemical or photoelectrochemical activities through the LEFIRC effect. As demonstrated previously, nanostructures at the laser processed surfaces change the state of water molecules through structural reorganization and wettability of surfaces.[49] Therefore, these surfaces can be used to modulate intermolecular and intramolecular interactions of the reagent molecules to further enhance the fuel generation rate.

**Supporting information**

Supporting Information related to this article can be found from the Opto-Electronic Advances or from the author.

**Acknowledgments**

We acknowledge financial supports from Jilin Province Science & Technology Development Project (20180414019GH), Bill & Melinda Gates Foundation (OPP1157723), National Key R&D Program of China (2018YFB1107202), National Natural Science Foundation of China (11774340), K.C. Wong Education Foundation (GJTD-2018-08), Scientific Research Project of the Chinese Academy of Sciences (QYZDB-SSW-SYS038) and Jilin Provincial Science & Technology Development Project (YDZJ202102CXJD002).

**Conflict of interest**

The authors declare no conflict of interest.

**Reference**

1. S. A. Montzka et al., "An unexpected and persistent increase in global emissions of ozone-depleting CFC-11," Nature **557**(7705), 413–417 (2018)

2. F. Knobloch et al., "Net emission reductions from electric cars and heat pumps in 59 world regions over time," Nature Sustainability **3**(6), 437–447 (2020).

3. C. Duan et al., "Highly efficient reversible protonic ceramic electrochemical cells for power generation and fuel production," Nature Energy **4**(3), 230–240 (2019).

4. T. Bak et al., "Photo-electrochemical hydrogen generation from water using solar energy.

Materials-related aspects," International Journal of Hydrogen Energy **27**(10), 991–1022 (2002).

5. C. S. Saraj et al., "Single-Step and Sustainable Fabrication of Ni(OH) 2 /Ni Foam Water Splitting Catalysts via Electric Field Assisted Pulsed Laser Ablation in Liquid," ChemElectroChem **8**(1), 209–217 (2021).

6. B. Lai et al., "Hydrogen evolution reaction from bare and surface-functionalized few-layered MoS2 nanosheets in acidic and alkaline electrolytes," Materials Today Chemistry **14**, 100207 (2019).

7. S. Shaik, S. P. de Visser, and D. Kumar, "External Electric Field Will Control the Selectivity of Enzymatic-Like Bond Activations," Journal of the American Chemical Society **126**(37), 11746–11749 (2004).

8. H. Hirao et al., "Effect of External Electric Fields on the C−H Bond Activation Reactivity of Nonheme Iron−Oxo Reagents," Journal of the American Chemical Society **130**(11), 3319–3327, American Chemical Society (2008).

9. Y.-J. Chung et al., "Coupling Effect of Piezo–Flexocatalytic Hydrogen Evolution with Hybrid 1T- and 2H-Phase Few-Layered MoSe2 Nanosheets," Advanced Energy Materials **10**(42), 2002082 (2020).

10. Y. Wu et al., "CO2 Reduction: Beyond d Orbits: Steering the Selectivity of Electrochemical CO2 Reduction via Hybridized sp Band of Sulfur-Incorporated Porous Cd Architectures with Dual Collaborative Sites (Adv. Energy Mater. 45/2020)," Advanced Energy Materials **10**(45), 2070183 (2020).

11. J. Z. Zhang and E. Reisner, "Advancing photosystem II photoelectrochemistry for semi-artificial photosynthesis," Nature Reviews Chemistry **4**(1), 6–21 (2020).

12. Y. Feng et al., "Self-powered electrochemical system by combining Fenton reaction and active chlorine generation for organic contaminant treatment," Nano Research **12**(11), 2729–2735 (2019).

13. S. Gao et al., "Self-Powered Electrochemical Oxidation of 4-Aminoazobenzene Driven by a Triboelectric Nanogenerator," ACS Nano **11**(1), 770–778, American Chemical Society (2017).

14. K. Mase et al., "Seawater usable for production and consumption of hydrogen peroxide as a solar fuel," Nature Communications **7**(1), 11470 (2016) [doi:10.1038/ncomms11470].

15. J. Xie et al., "Defect-Rich MoS 2 Ultrathin Nanosheets with Additional Active Edge Sites for Enhanced Electrocatalytic Hydrogen Evolution," Advanced Materials **25**(40), 5807–5813 (2013).

16. J. Benson et al., "Electrocatalytic Hydrogen Evolution Reaction on Edges of a Few Layer Molybdenum Disulfide Nanodots," ACS Applied Materials & Interfaces **7**(25), 14113–14122 (2015).

17. D. Kong et al., "Synthesis of MoS2 and MoSe2 Films with Vertically Aligned Layers," Nano Letters **13**(3), 1341–1347 (2013).

18. J. Kibsgaard et al., "Engineering the surface structure of MoS2 to preferentially expose active edge sites for electrocatalysis," Nature Materials **11**(11), 963–969 (2012).

19. M. Gong et al., "An Advanced Ni–Fe Layered Double Hydroxide Electrocatalyst for

Water Oxidation," Journal of the American Chemical Society **135**(23), 8452–8455 (2013).

20. Z. Chen et al., "Core–shell MoO3–MoS2 Nanowires for Hydrogen Evolution: A Functional Design for Electrocatalytic Materials," Nano Letters **11**(10), 4168–4175 (2011).

21. H. Zhang et al., "Small Dopants Make Big Differences: Enhanced Electrocatalytic Performance of MoS2 Monolayer for Oxygen Reduction Reaction (ORR) by N– and P– Doping," Electrochimica Acta **225**, 543–550 (2017).

22. H. Wang et al., "Transition-metal doped edge sites in vertically aligned MoS2 catalysts for enhanced hydrogen evolution," Nano Research **8**(2), 566–575 (2015).

23. C. Tsai, F. Abild-Pedersen, and J. K. Nørskov, "Tuning the MoS2 Edge-Site Activity for Hydrogen Evolution via Support Interactions," Nano Letters **14**(3), 1381–1387 (2014).

24. V. V. Welborn, L. Ruiz Pestana, and T. Head-Gordon, "Computational optimization of electric fields for better catalysis design," Nature Catalysis **1**(9), 649–655 (2018).

25. S. D. Fried, S. Bagchi, and S. G. Boxer, "Extreme electric fields power catalysis in the active site of ketosteroid isomerase," Science **346**(6216), 1510–1514, American Association for the Advancement of Science (2014).

26. M. Liu et al., "Enhanced electrocatalytic CO2 reduction via field-induced reagent concentration," Nature **537**(7620), 382–386 (2016).

27. T. Saberi Safaei et al., "High-Density Nanosharp Microstructures Enable Efficient CO2 Electroreduction," Nano Letters **16**(11), 7224–7228, (2016).

28. C. Geng et al., "Oriented external electric fields as mimics for probing the role of metal

ions and ligands in the thermal gas-phase activation of methane," Dalton Trans. **47**(43), 15271–15277, The Royal Society of Chemistry (2018).

29. S. Shaik, D. Mandal, and R. Ramanan, "Oriented electric fields as future smart reagents in chemistry," Nature Chemistry **8**(12), 1091–1098 (2016).

30. X. Huang et al., "Electric field induced selective catalysis of single-molecule reaction," Science Advances **5**(6), 3072, American Association for the Advancement of Science (2019).

31. Q.-X. Chen et al., "Ordered Nanostructure Enhances Electrocatalytic Performance by Directional Micro-Electric Field," Journal of the American Chemical Society **141**(27), 10729–10735 (2019).

32. J. Zhang et al., "Plasmonic metasurfaces with 42.3% transmission efficiency in the visible," Light: Science & Applications **8**(1), 53 (2019).

33. T. Bhuvana and G. U. Kulkarni, "Highly Conducting Patterned Pd Nanowires by Direct-Write Electron Beam Lithography," ACS Nano **2**(3), 457–462 (2008).

34. J. Zhang, C. Cong, and C. Guo, "Single-step maskless nano-lithography on glass by femtosecond laser processing," Journal of Applied Physics **127**(16), AIP Publishing LLC (2020).

35. Y. He et al., "Maskless laser nano-lithography of glass through sequential activation of multi-threshold ablation," Applied Physics Letters **114**(13) (2019).

36. M. Liu et al., "Enhanced electrocatalytic CO2 reduction via field-induced reagent concentration," Nature **537**(7620), 382–386, Nature Publishing Group (2016).

37. J. Bonse and S. Gräf, "Maxwell Meets Marangoni—A Review of Theories on Laser-Induced Periodic Surface Structures," Laser & Photonics Reviews **14**(10), 2000215 (2020).

38. J. Bonse et al., "Laser-Induced Periodic Surface Structures— A Scientific Evergreen," IEEE Journal of Selected Topics in Quantum Electronics **23**(3) (2017).

39. B. Wang et al., "Design of Aluminum Bowtie Nanoantenna Array with Geometrical Control to Tune LSPR from UV to Near-IR for Optical Sensing," Plasmonics **15**(3), 609–621 (2020).

40. S. K. Chamoli, S. C. Singh, and C. Guo, "1-D Metal-Dielectric-Metal Grating Structure as an Ultra-Narrowband Perfect Plasmonic Absorber in the Visible and Its Application in Glucose Detection," Plasmonics **15**(5), 1339–1350, Plasmonics (2020).

41. B. Wang et al., "Boosting Perovskite Photodetector Performance in NIR Using Plasmonic Bowtie Nanoantenna Arrays," Small **16**(24), 2001417 (2020).

42. C. Yao et al., "All-optical logic gates using dielectric-loaded waveguides with quasi-rhombus metasurfaces," Opt. Lett. **45**(13), 3769–3772, OSA (2020).

43. S. A. Jalil et al., "Multipronged heat-exchanger based on femtosecond laser-nano/microstructured Aluminum for thermoelectric heat scavengers," Nano Energy **75**, 104987 (2020).

44. B. Wang et al., "SERS study on the synergistic effects of electric field enhancement and charge transfer in an Ag2S quantum dots/plasmonic bowtie nanoantenna composite system," Photon. Res. **8**(4), 548–563, OSA (2020).

45. S. A. Jalil et al., "Formation of uniform two-dimensional subwavelength structures by delayed triple femtosecond laser pulse irradiation," Opt. Lett. **44**(9), 2278–2281, OSA (2019).

46. C. Yao et al., "Dielectric Nanoaperture Metasurfaces in Silicon Waveguides for Efficient and Broadband Mode Conversion with an Ultrasmall Footprint," Advanced Optical Materials **8**(17), 2000529 (2020).

47. S. K. Chamoli, S. Singh, and C. Guo, "Metal–Dielectric–Metal Metamaterial-Based Hydrogen Sensors in the Water Transmission Window," IEEE Sensors Letters **4**(5), 1–4 (2020).

48. S. K. Chamoli, S. C. Singh, and C. Guo, "Design of Extremely Sensitive Refractive Index Sensors in Infrared for Blood Glucose Detection," IEEE Sensors Journal **20**(9), 4628–4634 (2020).

49. S. C. Singh et al., "Solar-trackable super-wicking black metal panel for photothermal water sanitation," Nature Sustainability **3**(11), 938–946 (2020).